\documentclass{article}
\usepackage[utf8]{inputenc}
\usepackage{amssymb}
\usepackage{geometry}
\usepackage[usenames,dvipsnames]{color}
\usepackage{bbold}
\usepackage{slashed}
\usepackage[dvipsnames]{xcolor}
\usepackage{graphicx}
\usepackage{mathtools}
\usepackage{todonotes}
\usepackage{tabularx}
\usepackage{float}
\usepackage{wrapfig}
\usepackage{booktabs}
\usepackage{hyperref}
\usepackage[
backend=biber,
style=alphabetic,
sorting=ynt
]{biblatex}
\usepackage{comment}

\newcommand{\highlight}[2][yellow]{\mathchoice%
  {\colorbox{#1}{$\displaystyle#2$}}%
  {\colorbox{#1}{$\textstyle#2$}}%
  {\colorbox{#1}{$\scriptstyle#2$}}%
  {\colorbox{#1}{$\scriptscriptstyle#2$}}}%

\DeclarePairedDelimiter\ket{\lvert}{\rangle}
\DeclarePairedDelimiterX\braket[2]{\langle}{\rangle}{#1 \delimsize\vert #2}

\newcommand{\footremember}[2]{%
\footnote{#2}
\newcounter{#1}
\setcounter{#1}{\value{footnote}}%
}
\newcommand{\footrecall}[1]{%
\footnotemark[\value{#1}]%
}

\title{Quantum Arithmetic for Directly Embedded Arrays}
\date{}

\author{Alberto Manzano$^{a}$\footremember{email}{\textit{Corresponding authors:} \href{mailto:alberto.manzano.herrero@udc.es}{alberto.manzano.herrero@udc.es} \textit{and} \href{mailto:daniele.musso@cesga.es}{daniele.musso@cesga.es}  }\footremember{contributions}{\textit{These authors contribute equally to this work.}}, 
Daniele Musso$^{b}$\footrecall{email} \footrecall{contributions},
Álvaro Leitao$^{a}$,

Andrés Gómez$^{b}$

,\\
Carlos Vázquez$^{a}$,

Gustavo Ordóñez$^{c}$

and
María Rodríguez-Nogueiras$^{c}$.

}

\begin{document}

\maketitle
\begin{center}\it{
$^{a}$Department of Mathematics and CITIC, Universidade da Coruña, A Coruña, Spain\\
\vspace{15pt}
$^{b}$Galicia Supercomputing Center (CESGA),
Santiago de Compostela, Spain\\
\vspace{15pt}
$^{c}$Global Risk Analytics, HSBC, United Kingdom\\
}
\end{center}
\vspace{25pt}
\begin{abstract}
    We describe a general-purpose framework to design quantum algorithms relying upon an efficient handling of arrays. The corner-stone of the framework is the direct embedding of information into quantum amplitudes, thus avoiding the need to deal with square roots or encode the information in registers. We discuss the entire pipeline, from data loading to information extraction. Particular attention is devoted to the definition of an efficient tool-kit of quantum arithmetic operations on arrays. We comment on strong and weak points of the proposed manipulations, especially in relation to an effective exploitation of quantum parallelism. Eventually, we give explicit examples regarding the manipulation of generic oracles.
\end{abstract}

\newpage
\tableofcontents

\newpage
\section{Introduction, motivation and main results}
\label{intro}

Quantum hardware and software are still in their early days of development, thus the design of quantum algorithms typically focuses on low-level operations. Although one should always keep in mind the hardware limitations, especially when describing possible near-term implementations of quantum algorithms, it is convenient to pursue higher levels of abstraction. Apart from its long-term and algorithmic interest, a more abstract and standardized approach serves practical purposes too, for example that of making the benchmarking of quantum computer performances a more solid and transparent process. In turn, this helps pushing the research and the development in quantum computation at all levels.

\begin{figure}[h]
\centering
\includegraphics[width=0.5\textwidth]{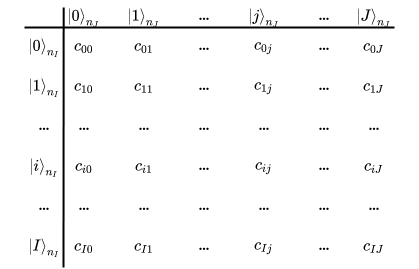}
\caption{Quantum matrix structure.}\label{fig:quantum_matrix}
\end{figure}

In the present paper, we describe a novel framework for the design of quantum algorithms on a more abstract plane. To this aim, our first proposal consists in the definition of a \emph{quantum matrix}, namely a quantum state organized in two registers: 
\begin{equation}\label{eqn:quantum_matrix_structure}
    \ket{\psi} = \sum_{i=0}^{I-1} \sum_{j=0}^{J-1}\, c_{ij}\, \ket{i}_{n_I}\otimes\ket{j}_{n_J}\ ,
\end{equation}
where $\ket{}_{n_J}$ indicates a register composed of $n_J$ qubits corresponding to $J = 2^{n_J}$ states, while $\ket{}_{n_I}$ is a register composed of $n_I$ qubits corresponding to $I = 2^{n_I}$ states. The overall state  $\ket{\psi}$, as defined in \eqref{eqn:quantum_matrix_structure}, is manifestly presented with the structure of a matrix; specifically, we interpret $i$ as the index running over the rows and $j$ as the index running over the columns. The rightmost qubit within a register is associated with the least significative digit of the associated index, in binary notation.%
\footnote{Thus, we are adopting a little-endian convention.} This way of storing the information has a common ground with that of Flexible Representation of Quantum Images (FRQI) and the Novel Enhanced Quantum Representation (NEQR) \cite{FRQI} \cite{NEQR}. The main difference with FRQI and NEQR is that we codify the information of the $(i,j)$ entry of the matrix in the quantum amplitude $c_{ij}$.%
 Intuitively, the matrix \eqref{eqn:quantum_matrix_structure} is a bi-dimensional memory array where $c_{ij}$ encodes the information stored in the $\ket{i}_{n_I}\otimes\ket{j}_{n_J}$ memory location (see Figure \ref{fig:quantum_matrix}).

The second proposal that we describe in the present work is a key technical feature about how we encode the information into the quantum amplitudes, the so-called \emph{direct embedding} \cite{kubo2020variational}.
Namely, the information to be stored into the quantum matrix is directly loaded into the amplitudes without taking square roots, as it is instead usually done in the literature.
Such loading choice has several and important implications in later stages of the quantum algorithms and --most importantly-- the information stored into the quantum state is handled and combined more easily, because algebraic operations are not hampered by the presence of square roots. This allows us to define an ``arithmetic library" composed of many fundamental arithmetic operations to handle arrays stored into the quantum matrix in an efficient manner. Such ``general purpose" library provides a versatile framework for the implementation of wide classes of algorithms. In this work, we provide some simple example algorithms without aiming to be exhaustive. The possibility of implementing arithmetic operations within a quantum framework has been considered in the literature since the early days of quantum computation. Apart from the quantum implementation of logical circuits corresponding to basic operations, like the quantum adder \cite{2000quantph8033D,compa}, also the manipulation of ``continuous" numbers has  been studied. Let us mention some works which, at least in spirit, are closer to ours \cite{wang2020quantum,PhysRevA.54.147,PhysRevLett.82.1784,QBLAS}. The difference with such approaches consists in the fact that we use a new embedding and organize the information into a matrix \eqref{eqn:quantum_matrix}; these two aspects combined allow us to work in a transparent and simple manner. For the same reason, the extraction of the information at the end of the quantum circuit requires strategies adapted to our encoding.

The third and final proposal in this paper is to give a full overview of the complete pipeline, or overall structure, of the generic algorithm admitting implementation within this framework. The first step of every algorithm corresponds to loading some input data. In the quantum case, it is often convenient to split this step into two sub-steps:
\begin{itemize}
    \item loading a probability distribution $p_{ij}$ \cite{grover_rudolph}\cite{nakaji2021approximate}\cite{kubo2020variational} 
    \item loading a bi-dimensional function $f_{ij}$ (possibly by means of methods that load information a line at a time).%
\footnote{In reference to the notation introduced for the quantum matrix \eqref{eqn:quantum_matrix_structure}, we have $c_{ij} = p_{ij} f_{ij}$, where the indexes are not summed over. 

}
\end{itemize}
It is not strictly necessary to split the loading into two steps. Yet, we consider such splitting because --typically-- we adopt different loading techniques for them: the probability distribution is loaded with a state preparation algorithm (\emph{e.g.} a multiplexor binary tree); the function is loaded by means of an auxiliary qubit meant to tell ``good" and ``bad" states apart. We describe the first step of the pipeline in Section \ref{sec:DataLoading}.

In Section \ref{sec:QuantumArithmetic} we describe the second step of the pipeline corresponding to the implementation of various arithmetic  operations, typically at the level of entire arrays or sub-arrays, and we refer to it as \emph{quantum arithmetic}. In Section \ref{sec:InformationExtraction} we describe the last step of the pipeline, which corresponds to extracting the information that we have stored in the quantum state, namely the read-out of the state that encodes the result of the algorithm. 
\begin{figure}[H]
\centering
\includegraphics[width=1\textwidth]{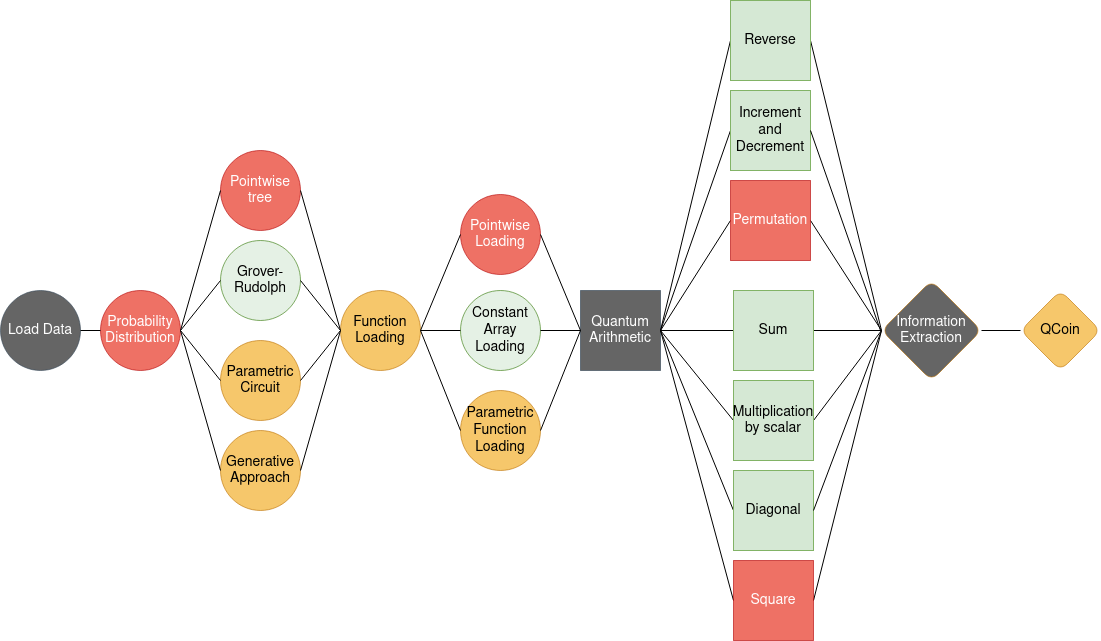}
\caption{Diagramatic structure of the pipeline. The dark grey nodes represent the three main steps; the nodes in red corresponds to algorithms that are not efficient, for those in yellow the efficiency varies from case to case and finally the nodes in green represent efficient algorithms.}\label{fig:pipeline}
\end{figure}
One of the advantages of organizing the pipeline as in Figure \ref{fig:pipeline} is that it enjoys a modular structure, therefore we can develop and analyse each of the steps independently, achieving a better understanding of the problems in each domain.
The color coding corresponds to the efficiency of the single modules.%
\footnote{It should be stressed that the efficiency of the single modules depicted in the diagram of Figure \ref{fig:pipeline} refers to the current state of the art which can change in the future, that is, it does not necessarily represent a structural limitation.} In particular, an overall efficient algorithm would correspond to an end-to-end green path from left to right across the diagram. In searching for possible implementations for a desired algorithm, the challenge is to improve the necessary blocks so to follow a completely green path.

\section{Data loading}\label{sec:DataLoading}

Data loading is a generic step which is required essentially in any quantum algorithm. The actual data to be loaded can vary in nature and serves different purposes. The data can correspond --for example-- to the discretization of a normalized general real function $f$ defined on a two-dimensional domain. This is, for instance, the typical setup needed for many tasks in mathematical finance, where the two dimensions represent an underlying price and time, respectively. Having in mind applications to finance, we focus on the loading of a real function upon a probability distribution. Yet, we can as well think of the loading of more general data corresponding to a complex matrix, as long as the normalization of the quantum state is respected.

The recipe described in the appendix works in a pointwise fashion, exploiting an auxiliary register to store the desired value into the quantum amplitude at each ``memory address", namely, to store it into the associated entry of the quantum matrix. It is important to underline from the outset that this pointwise approach is generically not efficient.
In order to attain  efficiency at the level of the full algorithm, we need to assume that the loading procedure can be implemented in an alternative and efficient way; in other words, we need to assume the existence of a suitable efficient oracle. Nonetheless, as we will show and stress later, a set of efficient manipulations for generic arrays is possible even when their loading is not efficient. This observation stems directly from the modular structure described in the pipeline of Figure \ref{fig:pipeline}.\\ \\

There are two different aspects related to the complexity of the state preparation, the quantum circuit complexity, on one side, and the complexity of the pre-processing algorithm (where needed), on the other. The former expresses a count of quantum operations or some quantitative estimation of the depth of the quantum loading circuit, the latter refers to the possible pre-processing needed to compute the case-specific values for the parameters of the quantum loading circuit.%
\footnote{An example of pre-processing algorithm would be the computation of the values of the angles in a tree-like loading (see for instance \cite{2002quant,Herbert_2021}).}
Here, we are going to discuss only the former, namely, the circuit complexity. To this purpose, we adopt the customary complexity indicator which simply relies on a count of the necessary CNOT gates.
This is motivated by the fact that CNOTs are sensibly more error-prone and require a longer execution time than single-qubit gates, as commented -for instance- in \cite{Shende_2006}.

Loading a generic real array is not  a trivial problem. In Appendix \ref{sec:PointwiseLoading} we referred to a pointwise loading, without worrying about its optimization.%
\footnote{Despite it not being optimal, we have adopted the pointwise method for its simplicity. 
We postpone the study of more optimized approaches to the future.} To this regard, the state of the art is currently set by two alternative approaches \cite{phdthesis}: one based on multiplexors \cite{2005quantph4100M,Shende_2006} and the other one based on Schmidt’s decomposition \cite{Plesch_2011,10.5555/2505466}. Both approaches give essentially the same leading CNOT complexity, namely, a number of CNOTs which scales as $2^{n+1}$ for the preparation of the generic $n$-qubit state.

Let us stress that, in the very specific case where we need to load a constant array, the procedure of Appendix \ref{sec:ConstantArray} requires (in the worst case scenario) $n_I$ $x$-gates, two $y$-rotations and one multi-controlled NOT gate. Such numbers must be compared with their classical counterpart, where the loading of a constant array on a line of the $I\times J$ matrix requires $J$ operations, considering that the process of copying a single number from memory is an operation.
Therefore, the loading of a constant function is more interesting from the quantum speed-up perspective than the pointwise loading of a generic function. Indeed, in principle, we need exponentially fewer operations on a quantum computer to load a constant array. Interestingly, the number of operations needed does \emph{not} depend on the length of the constant array that we want to load, but it does depend on the number of rows of the matrix that we have to control. Here we can directly see the nature of quantum systems in practice, there is an ``extra" cost associated to acting on a single element of the system without impacting the others. That makes operations on single elements inefficient and operations on the whole structure very efficient.

\section{Quantum arithmetic}\label{sec:QuantumArithmetic}

In the present section we provide a collection of tools for the efficient arithmetic handling of arrays encoded into a quantum matrix through direct embedding.
These tools have been implemented and tested using Qiskit \cite{Qiskit}. We present a selected set of implemented operations. Other operations potentially implementable within this framework are --for instance-- those described in \cite{diagonal}.

\subsection{Ordering}\label{sec:Ordering}

The first operations that we introduce are those which allow us to move elements within the quantum matrix. Manipulating single elements in the matrix has a much higher cost than performing operations on the whole structure. For this reason, we first introduce a global reversing operation and then we introduce generic permutations.
\subsubsection{Reversing}\label{sec:Reverse}
By reversing, we mean the operation
\begin{equation}
 R: (c_{i,0},c_{i,1},c_{i,2},c_{i,3},...,c_{i,J-1})\longrightarrow (c_{i,J-1},c_{i,J-2},c_{i,J-3},...,c_{i,0})\ ,
\end{equation}
where, for concreteness, we have addressed the reversing operation on the $i$-th row of the quantum matrix. Note that it is straightforward to perform the reversing operation on a column. 

We divide the process in three steps:
\begin{itemize}
    \item Mask the row. In this case we only need to mask the register corresponding to the row (the $\ket{}_{n_I}$ register) and leave the column register untouched. For more information on this operation see Appendix \ref{sec:PointwiseLoading}.
    \item Apply $n_J$ controlled $x$-gates. The controlled qubits are those of the row register. The target qubits are those of the column register.
    \item Undo step one, by applying again the same masking operation as before.
\end{itemize}
Following the steps above, we can perform a reversing operation on any row of the quantum matrix. 

If we wanted instead to reverse the whole matrix, the operation would be more efficient than just reversing a row or a column.%
\footnote{The inversion of the entire matrix corresponds to reading its entries from the lower-right corner \emph{i.e.} in the opposite direction to how matrix entries are usually read.}
In that case, there is no need to control on any qubit, we just need to apply an $x$-gate to each register of the quantum matrix.

As an explicit example, let us think of an $I\times J$ quantum matrix and,  for simplicity, let us consider $I = 2$. Hence, we need $n_I = 1$ and $n_J = \log_2(J)$. Suppose that we have loaded the following quantum matrix:
\begin{equation}
    \ket{\chi _1} = \dfrac{1}{A\sqrt{IJ}}\left(\sum _{j = 0}^{J-1}c_{0j}\ket{0}_{n_I}\otimes\ket{j}_{n_J}+\sum _{j = 0}^{J-1}c_{1j}\ket{1}_{n_I}\otimes\ket{j}_{n_J}\right).
\end{equation}
In order to reverse the first row, we start by applying an $x$-gate to the row register obtaining the state:
\begin{equation}
    \ket{\chi _2} = \dfrac{1}{A\sqrt{IJ}}\left(\sum _{j = 0}^{J-1}c_{0j}\ket{1}_{n_I}\otimes\ket{j}_{n_J}+\sum _{j = 0}^{J-1}c_{1j}\ket{0}_{n_I}\otimes\ket{j}_{n_J}\right).
\end{equation}
Now, the row on which we are focusing, namely the one corresponding to $c_{0j}$ for $j=0,...,J-1$, has all the qubits in the row register set to one (in this case the row register is just the qubit $\ket{}_{n_I}$). So, by means of controlled operations on $\ket{\chi_2}$, we act only on the row $c_{0j}$. Specifically, we apply an $x$-gate controlled on the row register, which acts on all the qubits of the column register. This yields
\begin{equation}
    \ket{\chi_3} = \dfrac{1}{A\sqrt{IJ}}\left(\sum _{j = 0}^{J-1}c_{0j}\ket{1}_{n_I}\otimes\ket{J-1-j}_{n_J}+\sum _{j = 0}^{J-1}c_{1j}\ket{0}_{n_I}\otimes\ket{j}_{n_J}\right).
\end{equation}
Finally, we apply again an $x$-gate to the row register obtaining:
\begin{equation}
    \ket{\chi _4} = \dfrac{1}{A\sqrt{IJ}}\left(\sum _{j = 0}^{J-1}c_{0j}\ket{0}_{n_I}\otimes\ket{J-1-j}_{n_J}+\sum _{j = 0}^{J-1}c_{1j}\ket{1}_{n_I}\otimes\ket{j}_{n_J}\right).
\end{equation}
The last step consists in undoing the mask.

\subsubsection{Permutations}
\label{sec:Permutations}

Permutations of two elements of an array, \emph{i.e.} swaps of two entries, are demanding operations as we have to manipulate individual elements, instead of whole blocks in the quantum matrix. 
For the sake of simplicity, in what follows we discuss the algorithm referring to a quantum matrix given by a single row. Generalizing to larger matrices is straightforward. It is relevant to point out that also the extension to higher dimensionalities, from bi-dimensional matrices to $d$-dimensional tensors, is doable, yet it requires additional controlled operations.%
\footnote{Note that the additional controlled operations may result in additional complexity.} Specifically, consider the state:
\begin{equation}
    \ket{\phi _1} = \dfrac{1}{\|f\|_{\infty}\, \sqrt{J}}\sum_{j = 0}^{J-1}f_j\ket{j}_{n_J}\ ,
\end{equation}
and let us write it in the notation:
\begin{equation}\label{eqn:array_permutation}
    (f_0,f_1,f_2,f_3,...,f_{J-4},f_{J-3},f_{J-2},f_{J-1}),
\end{equation}
which is more convenient to understand how the different gates act on the order of the components. The strategy presented here to perform a permutation of two arbitrary elements in \eqref{eqn:array_permutation} consists in using a pivot. That is, we choose a fixed position $k$ (pivot) and implement the permutations of the component placed at position $k$ and any other component in the array. Once this is done, the generic swap of two elements can be obtained by means of three operations, at most. For example, if we aim to make a permutation of elements in positions $i$ and $j$ in
\begin{equation}
    (f_0,...,f_i,...,f_k,...,f_j,...,f_{J-1})\ ,
\end{equation}
we would need to perform the following three steps: First, we permute the positions $j\Longleftrightarrow k$, obtaining
\begin{equation}
    (f_0,...,f_i,...,f_j,...,f_k,...,f_{J-1})\ .
\end{equation}
Then, we consider the permutation of positions $i\Longleftrightarrow k$ corresponding to the permutation of elements $i\Longleftrightarrow j$:
\begin{equation}
    (f_0,...,f_j,...,f_i,...,f_k,...,f_{J-1})\ .
\end{equation}
Finally, we perform again step one, obtaining the desired permuted state, namely
\begin{equation}
    (f_0,...,f_j,...,f_k,...,f_i,...,f_{J-1})\ .
\end{equation}

Now, the key in the algorithm is to understand how to actually perform in practice the permutations with the pivot. They can be implemented through $x$-gates and controlled $x$-gates. Moreover --without losing generality-- we choose the last element of the register as the pivot. If we have $n_J$ qubits, the single $x$-gates acting on state  \eqref{eqn:array_permutation} have the effects described in Table \ref{tab_1}.
\begin{table}[H]
\begin{center}
\resizebox{\textwidth}{!}{
\begin{tabular}{|c|c|c|}
\hline
 Gate & Old State & New State\\ 
 \hline
 $\mathbb{1}^{\otimes n_J-1}\otimes X$ & $f_0,f_1,f_2,f_3,...,f_{J-4},f_{J-3},\highlight[BurntOrange]{f_{J-2}},\highlight[yellow]{f_{J-1}})$ & $f_1,f_0,f_3,f_2,...,f_{J-3},f_{J-4},\highlight[yellow]{f_{J-1}},\highlight[BurntOrange]{f_{J-2}})$\\
 \hline
  $\mathbb{1}^{\otimes n_J-2}\otimes X \otimes \mathbb{1} $& $(f_0,f_1,f_2,f_3,...,\highlight[BurntOrange]{f_{J-4},f_{J-3}},\highlight[yellow]{f_{J-2},f_{J-1}})$ & $(f_2,f_3,f_0,f_1,...,\highlight[yellow]{f_{J-2},f_{J-1}},\highlight[BurntOrange]{f_{J-4},f_{J-3}})$\\
 \hline
  ... & ... & ...\\
 \hline
  $X\otimes \mathbb{1}^{\otimes n_J-1}$ & $(\highlight[BurntOrange]{f_0,f_1,f_2,f_3,...,f_{J/2-1}},\highlight[yellow]{f_{J/2},...,f_{J-4},f_{J-3},f_{J-2},f_{J-1}})$ & $(\highlight[yellow]{f_{J/2},f_{J/2+1},f_{J/2+2},f_{J/2+3},...,f_{J-1}},\highlight[BurntOrange]{f_{0},...,f_{J/2-4},f_{J/2-3},f_{J/2-2},f_{J/2-1}})$\\
  \hline
\end{tabular}
}
\end{center}
\caption{Note how each gate affects the whole state, however the coloring emphasizes the effect of the gates on the rightmost blocks.
}
\label{tab_1}
\end{table}

From Table \ref{tab_1} we can see that the single $x$-gates are performing swaps of blocks of \textit{contiguous memory positions}. When we act on more significant qubits we are affecting bigger blocks and each gate is affecting the whole state. In this algorithm we are only interested on the effect that the gate has on certain blocks of the array (the ones highlighted). Using multi-controlled $x$-gates where the controls are applied to all qubits (except the one where we apply the $x$ gate) and acting on state \eqref{eqn:array_permutation}, we get the results reported in Table \ref{tab_2}.
\begin{table}[H]
\begin{center}
\resizebox{\textwidth}{!}{
\begin{tabular}{|c|c|c|}
\hline
 Gate & Old State & New State\\ 
 \hline
 $c^{\otimes n_J-1}\otimes X$ & $(f_0,f_1,f_2,f_3,...,f_{J-4},f_{J-3},\highlight[BurntOrange]{f_{J-2}},\highlight[yellow]{f_{J-1}})$ & $(f_0,f_1,f_2,f_3,...,f_{J-4},f_{J-3},\highlight[yellow]{f_{J-1}},\highlight[BurntOrange]{f_{J-2}})$\\
 \hline
  $c^{\otimes n_J-2}\otimes X \otimes c $ & $(f_0,f_1,f_2,f_3,...,\highlight[BurntOrange]{f_{J-4},f_{J-3}},\highlight[yellow]{f_{J-2},f_{J-1}})$ & $(f_0,f_1,f_2,f_3,...,f_{J-4},\highlight[yellow]{f_{J-1}},f_{J-2},\highlight[BurntOrange]{f_{J-3}})$\\
 \hline
  ... & ... & ...\\
 \hline
  $X\otimes c^{\otimes n_J-1}$ & $(\highlight[BurntOrange]{f_0,f_1,f_2,f_3,...,f_{J/2-1}},\highlight[yellow]{f_{J/2},...,f_{J-4},f_{J-3},f_{J-2},f_{J-1}})$ & $(f_0,f_1,f_2,f_3,...,\highlight[yellow]{f_{J-1}},f_{J/2},...,f_{J-4},f_{J-3},f_{J-2},\highlight[Orange]{f_{J/2-1}})$\\
  \hline
\end{tabular}
}
\end{center}
\caption{The use of controlled gates has two different effects. In the second column of the table, we are emphasizing that they connect certain blocks of the matrix (those highlighted). In the third column of the table, we are emphasizing the fact that only the rightmost entries of each block are impacted.}
\label{tab_2}
\end{table} \noindent
In this case it is clear that the effect of the controlled operations is to permute the last elements of each highlighted block.

We need to combine both operations, $x$-gates and multi-controlled $x$-gates, to perform the permutation of any element with the pivot (the last element, according to our choice). The strategy can be implemented recursively in the following way:
\begin{enumerate}
    \item Move the last element of the array to the block where the element we wish to permute is located. This is done through a suitable multi-controlled $x$-gate.
    \item If at this point the two elements that we wanted to interchange have been actually swapped, then undo all previous operations (both $x$-gates and multi-controlled $x$-gates) except for the last one and finish. These operations are needed to bring back to their original position all the other elements except the pair that has been swapped.
    Otherwise continue.
    \item Swap the blocks on which we have acted at step 1. This is done through a single $x$-gate and serves the purpose of moving to the right the block on which we need to focus.
    \item Go back to step 1.
\end{enumerate}

For the sake of clarity, let us give a simple explicit example. Consider the state
\begin{equation}
    \ket{\psi} = \left(f_0,f_1,f_2,f_3,f_4,f_5,f_6,f_7\right)\ ,
\end{equation}
and suppose we want to permute the first element $f_0$ with the pivot element $f_7$. We can proceed as follows:
\begin{align*}
    \left(X\otimes c\otimes c\right)\left(\highlight[BurntOrange]{f_0,f_1,f_2,f_3},\highlight[yellow]{f_4,f_5,f_6,f_7}\right)&=\left(f_0,f_1,f_2,\highlight[yellow]{f_7},f_4,f_5,f_6,\highlight[BurntOrange]{f_3}\right) ,\\
    \left( X\otimes \mathbb 1 \otimes  \mathbb 1\right)\left(\highlight[BurntOrange]{f_0,f_1,f_2,f_7},\highlight[yellow]{f_4,f_5,f_6,f_3}\right)&=\left(\highlight[yellow]{f_4,f_5,f_6,f_3},\highlight[BurntOrange]{f_0,f_1,f_2,f_7}\right) ,\\
    \left(c\otimes X\otimes c\right)\left(f_4,f_5,f_6,f_3,\highlight[BurntOrange]{f_0,f_1},\highlight[yellow]{f_2,f_7}\right)&=\left(f_4,f_5,f_6,f_3,f_0,\highlight[yellow]{f_7},f_2,\highlight[BurntOrange]{f_1}\right),\\
    \left(\mathbb 1\otimes X\otimes \mathbb 1\right)\left(f_4,f_5,f_6,f_3,\highlight[BurntOrange]{f_0,f_7},\highlight[yellow]{f_2,f_1}\right)&=\left(f_6,f_3,f_4,f_5,\highlight[yellow]{f_2,f_1},\highlight[BurntOrange]{f_0,f_7}\right),\\
    \left(c\otimes c\otimes X\right)\left(f_6,f_3,f_4,f_5,f_2,f_1,\highlight[BurntOrange]{f_0},\highlight[yellow]{f_7}\right)&=\left(f_6,f_3,f_4,f_5,f_2,f_1,\highlight[yellow]{f_7},\highlight[BurntOrange]{f_0}\right).\\
    \end{align*}
    Now that we have effectively swapped the element $0$ and the element $7$ we just have to relocate the rest of the elements (Step 3).
    \begin{align*}
    \left(\mathbb 1\otimes X\otimes \mathbb 1\right)\left(f_6,f_3,f_4,f_5,\highlight[BurntOrange]{f_2,f_1},\highlight[yellow]{f_7,f_0}\right)&=,\left(f_4,f_5,f_6,f_3,\highlight[yellow]{f_7,f_0},\highlight[BurntOrange]{f_2,f_1}\right),\\
    \left(c\otimes X\otimes c\right)\left(f_4,f_5,f_6,f_3,\highlight[BurntOrange]{f_7,f_0},\highlight[yellow]{f_2,f_1}\right)&=\left(f_4,f_5,f_6,f_3,f_7,\highlight[yellow]{f_1},f_2,\highlight[BurntOrange]{f_0}\right),\\
    \left( X\otimes \mathbb 1 \otimes  \mathbb 1\right)\left(\highlight[BurntOrange]{f_4,f_5,f_6,f_3},\highlight[yellow]{f_7,f_1,f_2,f_0}\right)&=\left(\highlight[yellow]{f_7,f_1,f_2,f_0},\highlight[BurntOrange]{f_4,f_5,f_6,f_3}\right),\\
    \left(X\otimes c\otimes c\right)\left(\highlight[BurntOrange]{f_7,f_1,f_2,f_0},\highlight[yellow]{f_4,f_5,f_6,f_3}\right)&=\left(f_7,f_1,f_2,\highlight[yellow]{f_3},f_4,f_5,f_6,\highlight[BurntOrange]{f_0}\right).\\
\end{align*}

\subsubsection{Cyclic permutations}

Cyclic permutations corresponds to the two transformations:
\begin{flalign}
 & \text{Left}: (f_{0},f_{1},f_{2},...,f_{J-2},f_{J-1})\longrightarrow (f_{1},f_{2},f_{3},...,f_{J-1},f_{0})\ ,\\
 & \text{Right}: (f_{0},f_{1},f_{2},...,f_{J-2},f_{J-1})\longrightarrow (f_{J-1},f_{0},f_{1},...,f_{J-3},f_{J-2})\ ,
\end{flalign}
where we follow the same notation adopted in Section \ref{sec:Permutations}.%
\footnote{In what follows, we adopt the notation $\supset$ to consider just some specific terms that are relevant for our purposes within a bigger quantum state.} These operators have been discussed in depth in \cite{cyclic_permutations} and their implementation can be immediately extended to our framework, upon adding suitable controls.

\subsection{Addition}
In this subsection we discuss both the sum of whole arrays and the sum of their components (reduction).
\subsubsection{Sum}\label{sec:Sum}

Consider the state $|\psi_1\rangle$ given in \eqref{eqn:pointwise_function_loaded}:
\begin{equation}\label{start}
 \begin{split}
    \ket{\psi_1} \supset &\dfrac{1}{\|f\|_{\infty}\, \sqrt{IJ}}\sum_{j = 0}^{J-1}\sum_{i = 0}^{I-1}f_{ij} \ket{i}_{n_I}\otimes\ket{j}_{n_J}\ ,
 \end{split}
\end{equation}
where we have omitted the auxiliary register $\ket{}_a$ for convenience. Applying a Hadamard gate on the first qubit of the row register, namely
\begin{equation}
 \ket{\psi_2} = \left(\mathbb{1}^{\otimes n_I-1}\otimes H\otimes \mathbb{1}^{\otimes n_J}\right) \ket{\psi_1}\ ,    
\end{equation}
we get the sum and difference of the the rows grouped in pairs, explicitly
\begin{equation}\label{eqn:sum}
 \begin{split}
    \ket{\psi_2} \supset &\dfrac{1}{\sqrt{2}\|f\|_{\infty}\, \sqrt{IJ}}\sum_{j = 0}^{J-1}\sum_{i = 0}^{\frac{I}{2}-1}\Bigg[\left(f_{2i,j}+f_{2i+1,j}\right) \ket{2i}_{n_I}\otimes\ket{j}_{n_J}
    \\&\qquad \qquad \qquad \qquad \qquad \qquad \qquad
    +\left(f_{2i,j}-f_{2i+1,j}\right) \ket{2i+1}_{n_I}\otimes\ket{j}_{n_J}\Bigg].\\
 \end{split}
\end{equation}
\color{black}
In the first row of \eqref{eqn:sum}, we get the sum of the first and second row of \eqref{start}. In the second row of \eqref{eqn:sum}, instead, we get the difference between the first and the second row of \eqref{start}. In the third row of \eqref{eqn:sum}, we get the sum of the third and fourth row of \eqref{start}, and the same structure continues on.

An analogous sum/difference operation can be performed in columns. Note that, in order to consider the correct number of $\frac{1}{\sqrt{2}}$ factors, we need to count the Hadamard gates that we apply.
Eventually, to sum two rows that are not in the same pair, we can take advantage of the permutations described in Section \ref{sec:Permutations}.

\subsubsection{Reductions}\label{sec:Reduction}
By \emph{reduction} we mean the summation of all the elements of an array
\begin{equation}\label{red_def}
 \text{Reduction}: (f_0,f_1,...,f_{J-1})\longrightarrow(f_0+f_1+...+f_{J-1},...)\ ,
\end{equation}
where the result of the reduction is stored into the first entry of the array. Consider again the state defined in \eqref{start}, that is, 
\begin{equation}
 \begin{split}
    \ket{\psi_1} \supset &\dfrac{1}{\|f\|_{\infty}\, \sqrt{IJ}}\sum_{j = 0}^{J-1}\sum_{i = 0}^{I-1}f_{ij} \ket{i}_{n_I}\otimes\ket{j}_{n_J}\ .
 \end{split}
\end{equation}
In order to perform a reduction by rows (\emph{i.e.} summing the elements of each row and storing the result on the first column), we just need to apply a Hadamard gate to every qubit of the column register 
\begin{equation}\label{WH}
 \ket{\psi_2} = \left(\mathbb{1}^{\otimes n_I}\otimes H^{\otimes n_J}\right)\ket{\psi _1}\ ,
\end{equation}
which gives:
\begin{equation}\label{eqn:reduction}
 \begin{split}
    \ket{\psi_2} \supset &\dfrac{1}{\sqrt{2}\|f\|_{\infty}\, \sqrt{IJ}}\sum_{i = 0}^{I-1}\Bigg(\sum_{j = 0}^{J-1}f_{ij}\Bigg)\ket{i}_{n_I}\otimes\ket{0}_{n_J}.\\
 \end{split}
\end{equation}
The parenthesis implies that we have the reduction of each row in the first column (which corresponds to $\ket{0}_{n_J}$). In the rest of the columns, we get other reductions with different combinations of signs, as implemented by the Walsh-Hadamard operator in \eqref{WH}.

If we were to do a reduction by columns, instead of by rows, we will need to apply the Walsh-Hadamard gate to the row register, instead of the column register. Correspondingly, we will get the reduction of the columns on the first row.

\subsection{Products}
In this subsection we consider the product of a whole array by a constant and the product of two arrays. Eventually, the scalar product of two arrays can be obtained composing the product of two arrays and a reduction. Similarly, the squaring of an array can be obtained as the product of the array by itself.

\subsubsection{Multiplication by a constant}

In order to multiply a row or a column by a constant, we need an extra qubit which we denote $\ket{0}_{\text{mul}}$. Consider the state defined in \eqref{eqn:pointwise_function_loaded}, but this time supplemented with the extra qubit: 
\begin{equation}
 \begin{split}
    \ket{\psi_1} \supset &\dfrac{1}{\|f\|_{\infty}\, \sqrt{IJ}}\ \ket{0}_{\text{mul}} \otimes \sum_{j = 0}^{J-1}\sum_{i = 0}^{I-1}f_{ij} \ket{i}_{n_I}\otimes\ket{j}_{n_J}\ .
 \end{split}
\end{equation}
The multiplication operation consists merely in a controlled rotation. The rotation needs to be applied onto the auxiliary register $\ket{0}_{\text{mul}}$, it introduces a factor $\alpha = \cos{\frac{\theta}{2}}$, so we are initially restricted to multiplication by numbers between $0$ and $1$. This limitation can be circumvented by means of suitable manipulations of the normalization constants.

Depending on the controls that we apply, we can multiply a row, a column or a specific individual entry by $\alpha$. For example, assume that we want to multiply the first row by $\alpha$. In order to act solely on the first row, we first have to mask it by applying
\begin{equation}
    \ket{\psi_2} = \left(X^{n_I}\otimes \mathbb{1}^{n_J}\right)\ket{\psi_1}\ ;
\end{equation}
explicitly, we obtain:
\begin{equation}
 \begin{split}
    \ket{\psi_2} \supset &\dfrac{1}{\|f\|_{\infty}\, \sqrt{IJ}}\ \ket{0}_{\text{mul}}\otimes \sum_{j = 0}^{J-1} f_{0j} \ket{I-1}_{n_I}\otimes\ket{j}_{n_J}+... \ .
 \end{split}
\end{equation}
The next step is to perform the controlled $y$-rotation
\begin{equation}\label{mul_con}
 \ket{\psi_3} = \left[\mathcal{R}_y(\theta)\otimes C^{\otimes n_I}\otimes \mathbb{1}^{\otimes n_J}\right] \ket{\psi_2}\ ,    
\end{equation}
where we have indicated the controls of the controlled rotations with the symbol $C$. Thus, \eqref{mul_con} is to be interpreted as a controlled $y$-rotation acting on $\ket{0}_{\text{mul}}$ and controlled by the row register.
Explicitly, \eqref{mul_con} gives
\begin{equation}
 \begin{split}
    \ket{\psi_3} \supset &\dfrac{1}{\|f\|_{\infty}\, \sqrt{IJ}}\cos\left(\dfrac{\theta}{2}\right)\ket{0}_{\text{mul}}\otimes\sum_{j = 0}^{J-1}f_{0j} \ket{I-1}_{n_I}\otimes\ket{j}_{n_J}\\
    +&\dfrac{1}{\|f\|_{\infty}\, \sqrt{IJ}}\sin\left(\dfrac{\theta}{2}\right)\ket{1}_{\text{mul}}\otimes\sum_{j = 0}^{J-1}f_{0j} \ket{I-1}_{n_I}\otimes\ket{j}_{n_J}+...\ .
 \end{split}
\end{equation}
Eventually, we have to unmask the state
\begin{equation}
 \ket{\psi_4} = \left(X^{n_I}\otimes \mathbb{1}^{n_J}\right)\ket{\psi_3}\ ,
\end{equation}
which yields
\begin{equation}
 \begin{split}
    \ket{\psi_4} \supset &\dfrac{1}{\|f\|_{\infty}\, \sqrt{IJ}}\cos\left(\dfrac{\theta}{2}\right)\ket{0}_{\text{mul}}\otimes \sum_{j = 0}^{J-1}f_{0j} \ket{0}_{n_I}\otimes\ket{j}_{n_J}\\
    &+\dfrac{1}{\|f\|_{\infty}\, \sqrt{IJ}}\sin\left(\dfrac{\theta}{2}\right)\ket{1}_{\text{mul}}\otimes \sum_{j = 0}^{J-1}f_{0j} \ket{0}_{n_I}\otimes\ket{j}_{n_J}\\
    &+\dfrac{1}{\|f\|_{\infty}\, \sqrt{IJ}}\ket{0}_{\text{mul}}\otimes \sum_{j = 0}^{J-1}\sum_{i = 1}^{I-1}f_{ij} \ket{i}_{n_I}\otimes\ket{j}_{n_J}\ .
 \end{split}
\end{equation}
The relevant information is marked by $\ket{0}_{\text{mul}}$.

\subsubsection{Array multiplication}\label{sec:product}

In the present section, we describe the theoretical proposal for a more advanced operation: the multiplication of arrays. Its (overall) efficiency is related to that of the loading process. Let us assume to dispose of the following oracles:
\begin{align}
 O_f\big(\ket{0}\otimes\ket{j}\big) &= f_j \ket{0}\otimes\ket{j} + \sqrt{1-f_j^2}\, \ket{1}\otimes\ket{j},\\
 O_g\big(\ket{0}\otimes\ket{j}\big) &= g_j \ket{0}\otimes\ket{j} + \sqrt{1-g_j^2}\, \ket{1}\otimes\ket{j},
\end{align}
which load the arrays $f$ and $g$ respectively. Morever, consider the swap operator:
\begin{equation}
    S\big(\ket{\alpha}\otimes\ket{\beta}\otimes\ket{j}\big) =
    \ket{\beta}\otimes\ket{\alpha}\otimes\ket{j}.
\end{equation}
To build the multiplication operator we are going to start from the state:
\begin{equation}
    \ket{\lambda_0} = \ket{0}_g\otimes\ket{0}_f\otimes\ket{j},
\end{equation}
where we have two auxiliary qubits: $\ket{0}_f$ to load $f$ and $ \ket{0}_g$ to load $g$. First we load $f$:
\begin{equation}
    \ket{\lambda_1} = \big(\mathbb{1}\otimes O_f\big)\ket{\lambda_0} = 
    f_j \ket{0}_g\otimes\ket{0}_f\otimes\ket{j}+ \sqrt{1-f_j^2}\, \ket{0}_g\otimes\ket{1}_f\otimes\ket{j}
\end{equation}
In the second step we swap qubits $\ket{}_f$ and $\ket{}_g$:
\begin{equation}
    \ket{\lambda_2} = S\ket{\lambda_1} = 
    f_j \ket{0}_f\otimes\ket{0}_g\otimes\ket{j} + \sqrt{1-f_j^2}\, \ket{1}_f\otimes\ket{0}_g\otimes\ket{j}.
\end{equation}
The third and last step consists in applying the oracle $O_g$:
\begin{align}
    \ket{\lambda_3} = \big(\mathbb{1}\otimes O_g\big)\ket{\lambda_2} &= 
    g_j f_j \ket{0}_f\otimes\ket{0}_g\otimes\ket{j} + g_j\, \sqrt{1-f_j^2}\, \ket{1}_f\otimes\ket{0}_g\otimes\ket{j} \\ \nonumber
    &\qquad \qquad + \sqrt{1-g_j^2}\, f_j \ket{0}_f\otimes\ket{1}_g\otimes\ket{j} + \sqrt{1-g_j^2}\,\sqrt{1-f_j^2}\, \ket{1}_f\otimes\ket{1}_g\otimes\ket{j}.
\end{align}
The multiplication of arrays $f$ and $g$ is encoded in the registers marked by $\ket{0}_f\otimes \ket{0}_g$:%
\footnote{In order to return back to the original ordering of the auxiliary qubits, $\ket{0}_g\otimes\ket{0}_f$, one can consider an extra swap $S$.}
\begin{equation}
    \ket{\lambda_3} =
    \big(\mathbb{1}\otimes O_g\big)\, S\, \big(\mathbb{1}\otimes O_f\big)
    \supset 
    g_j f_j \ket{0}_f\otimes\ket{0}_g\otimes\ket{j}\ .
\end{equation}
This procedure, when extended to the multiplication of more than two arrays. It is worth noticing the fact that this method depends on the loading complexity, that is, the efficiency of the employed oracles.

As a final comment, when in Section \ref{intro} we split the loading of the integrand $f\cdot p$ in two parts, that associated to the distribution and that associated to the function, we were in fact performing the multiplication of the two arrays $\{p_j\}$ and $\{f_j\}$.

\subsubsection{Squaring and scalar product}\label{sec:ScalarProduct}

The square of an array and the scalar product of two arrays can be obtained from operations that we have already defined above. The former is trivially just the multiplication of an array by itself. On the other hand, if we perform the reduction of the product of two arrays, we get their scalar product. As their construction depends on the steps commented in Section \ref{sec:product}, the efficiency of the square of an array and the scalar product of two arrays is strongly dependent on the loading strategy for the arrays.

\section{Information extraction}\label{sec:InformationExtraction}

In this section we describe a method to extract the information stored in the quantum circuit at the end of the algorithm. The method is called Quantum Coin \cite{Abrams1999FastQA,shimada2020quantum} (QCoin) and it can be applied in general to extract the absolute value of the amplitude $a$ along a state $\ket{\chi}$. Although in this article we rely on the QCoin algorithm, there are other techniques which can be considered, such as those appearing in \cite{AAE}\cite{Suzuki_2020}\cite{grinko2020iterative} \cite{giurgicatiron2020low}.%
\footnote{Another information extraction method compatible with the direct embedding has been described in \cite{kubo2020variational}; such method is tailored to compute expectation values of functions admitting a polynomial approximation.} 

Let us suppose that we want to extract the amplitude of the state
\begin{equation}
    \ket{\chi} = 
    \ket{1}_a\otimes\ket{i}_{n_I}\otimes\ket{j}_{n_J}\ ,
\end{equation}
for some specific values $i$ and $j$. For example, we want to read one entry of the quantum matrix defined in \eqref{eqn:quantum_matrix_structure}. 
The QCoin algorithm starts with an unamplified estimation of the amplitude along $\ket{\chi}$, which is then iteratively refined. Such unamplified estimation consists in a repeated set of identical experiments and measurements, which allows us to get an empirical value $\tilde \mu$ for the absolute value of the desired amplitude. This estimation provides a confidence interval 
\begin{equation}\label{conf_int}
    [\tilde\mu-\delta,\tilde\mu+\delta]\ ,
\end{equation} 
relying on standard statistical tools, \emph{e.g.} the Chebyshev's inequality.
Then, one can pursue an iterative Grover amplification strategy,
implemented as a ``zoom-in" operation into the confidence interval estimated previously. This amounts to an iterative exponential refinement of the accuracy of the estimation of $\tilde \mu$ \cite{Abrams1999FastQA}. We give some additional details below.

\subsection{Amplified iterations}
\label{sub_ext}

An amplified stage consists of two operations:
\begin{itemize}
 \item 
 A constant shift of the amplitude $a$ by the position of the lower bound of the confidence interval \eqref{conf_int} of its empirically estimated value, namely 
 \begin{equation}\label{shi}
     a\rightarrow \bar a = a - (\tilde\mu -\delta)\ .
 \end{equation}
 By construction, the shifted amplitude $\bar a$ is expected to have a positive value of the order of half the size of the confidence interval.%
 \footnote{It is crucial to consider circumstances where the expected value is positive because the algorithm is insensitive to the sign. Similarly, note that --despite the loading module and the arithmetic manipulation module can be straightforwardly extended to complex amplitudes-- the read-out would need to be phase sensitive.}
 \item 
 An enhancement of the probability $P$ of measuring $|\chi\rangle$. Specifically, by means of Grover amplification, we obtain 
 a state where the probability $\hat P$ of getting $\ket{\chi}$ is 
 \begin{equation}\label{enh}
  \hat P = \gamma P\ ,
 \end{equation}
with $\gamma > 1$. An estimation of $\hat P$ with precision $\epsilon$ corresponds to an estimation 
of $P$ with increased precision $\frac{\epsilon}{\gamma}$. 
\end{itemize}

We define the Grover amplifier as usual
\begin{equation}
 Q = O_{\bar a}\, R_{\sigma}\, O_{\bar a}^{-1}\, R_{\chi}\ ,
\end{equation}
where $\sigma$ is the state
\begin{equation}
    |\sigma\rangle \equiv \ket{1}_a\otimes\ket{0}_{n_I}\otimes\ket{0}_{n_J}\ ,
\end{equation}
and where $O_{\bar a}$ represents the operator which implements the algorithm followed by the shift \eqref{shi}.
Note that the state $\ket{\sigma}$ is the entry marked as $00$ in the quantum matrix \eqref{eqn:quantum_matrix}. We denote with $R_{\sigma}$ and $R_{\chi}$ the reflection operators given respectively by
\begin{equation}
 R_{\sigma} = \mathbb 1 - 2 |\sigma\rangle\langle \sigma|\ ,\quad 
 R_{\chi} = \mathbb 1 - 2 |\chi\rangle\langle \chi|\ .
\end{equation}
If we define the angle $\theta$, such that
\begin{equation}
 \sin(\theta) \equiv \langle \chi|\Psi\rangle\ ,
\end{equation}
where $\ket{\Psi}$ is the final quantum state, we have $P=\sin(\theta)^2$ and
\begin{equation}
 \langle \chi | Q^k |\Psi\rangle = \sin[(2k+1)\theta]\ .
\end{equation}
The corresponding amplification factor $\gamma$ for the probability of measuring $|\chi\rangle$ is
\begin{equation}
 \gamma = \frac{\sin[(2k+1)\theta]^2}{\sin(\theta)^2}\ .
\end{equation}
In order to maximize $\gamma$ we need to choose $k$ so that
\begin{equation}
 (2k+1) \theta \lesssim \frac{\pi}{2}\ .
\end{equation}

\section{Examples}

In this section we give two explicit examples of manipulations for a given oracle.

\subsection{Constant shift of a given oracle}
\label{shift}

Given an oracle $O_f$ which loads the function $f$, we are interested in providing an efficient implementation of an oracle $O_{\bar f}$ for the shifted function 
\begin{equation}
 \bar f_j = f_j - s\ ,
\end{equation}
where $s$ is a generic real constant. 
Note that constructing the oracle for $\bar f$ is non-trivial whenever $f$ is non-constant.%
\footnote{The constant shift of a constant function amounts to just a global rotation. However, the constant shift of a non-constant function requires in general a position-dependent rotation.}

More specifically, consider the state
\begin{align}\label{fun_row}
 \ket{\phi_1} = &\dfrac{1}{\|f\|_{\infty}\, \sqrt{2J}}\sum_{j = 0}^{J-1}\Bigg(f_{j} \ket{0}_a\otimes\ket{0}_1\otimes\ket{j}_{n_J}+\sqrt{1-f_{j}^2} \ket{1}_a\otimes\ket{0}_1\otimes\ket{j}_{n_I}\Bigg)\\ \nonumber
 &+\dfrac{1}{\sqrt{2J}}\sum_{j = 0}^{J-1}\Bigg( \ket{0}_a\otimes\ket{1}_1\otimes\ket{j}_{n_J}+ \ket{1}_a\otimes\ket{1}_1\otimes\ket{j}_{n_J}\Bigg)\ ,
\end{align}
which is the result of applying the oracle $O_f$ to the initial state $|\psi_0\rangle$ given in \eqref{eqn:uniform_distribution_loaded}. Note that the state $\ket{\phi_1}$ can be interpreted as having loaded the function $f$ on the first row of the $2\times J$ quantum matrix.

Following the steps described in Appendix \ref{sec:ConstantArray}, we can load the constant function $s$ into the second row of $|\phi_1\rangle$. Specifically, we load a constant array on the states associated to $i=1$. It is important to remember that the quantum register $|i\rangle_{n_I}$ stores the row address of the arrays, while $|j\rangle_{n_J}$ stores the column address.

More explicitly, we obtain
\begin{align}
 \ket{\phi_2} = &\dfrac{1}{\|f\|_{\infty}\, \sqrt{2J}}\sum_{j = 0}^{J-1}\Bigg(f_{j} \ket{0}_a\otimes\ket{0}_1\otimes\ket{j}_{n_J}+\sqrt{1-f_{j}^2} \ket{1}_a\otimes\ket{0}_1\otimes\ket{j}_{n_I}\Bigg)\\ \nonumber
 &+\dfrac{1}{\sqrt{2J}}\sum_{j = 0}^{J-1}\Bigg( s\ket{0}_a\otimes\ket{1}_1\otimes\ket{j}_{n_J}+ \sqrt{1-s^2}\ket{1}_a\otimes\ket{1}_1\otimes\ket{j}_{n_J}\Bigg)\ .
\end{align}

Following Section \ref{sec:Sum}, we apply a Hadamard gate to the row qubit combining the two rows of the stored matrix. Namely
\begin{equation}\label{psi_3}
 \begin{split}
    \ket{\phi_3} &=\left(\mathbb 1 \otimes H\otimes \mathbb 1^{\otimes n_J}\right) \ket{\phi_2}\\ &=\dfrac{1}{2\|f\|_{\infty}\sqrt{J}}\sum_{j = 0}^{J-1}\Bigg(
    f_j \ket{0}_a\otimes\ket{0}_1\otimes\ket{j}_{n_J}
    +f_j \ket{0}_a\otimes\ket{1}_1\otimes\ket{j}_{n_J}\\
    &+\sqrt{1-f_j^2} \ket{1}_a\otimes\ket{0}_1\otimes\ket{j}_{n_J}
    +\sqrt{1-f_j^2} \ket{1}_a\otimes\ket{1}_1\otimes\ket{j}_{n_J}\\
    &+s \ket{0}_a\otimes\ket{0}_1\otimes\ket{j}_{n_J}
    -s \ket{0}_a\otimes\ket{1}_1\otimes\ket{j}_{n_J}\\
    &+\sqrt{1-s^2} \ket{1}_a\otimes\ket{0}_1\otimes\ket{j}_{n_J}
    -\sqrt{1-s^2} \ket{1}_a\otimes\ket{1}_1\otimes\ket{j}_{n_J}\Bigg)\ .
 \end{split}
\end{equation}
Let us focus just on the components $|0\rangle_a \otimes |1\rangle\otimes |j\rangle_{n_J}$, namely
\begin{equation}\label{diff}
    \ket{\phi_3} \supset \ket{\phi_3^{(-)}} \equiv \dfrac{1}{2\|f\|_{\infty}\sqrt{J}}\sum_{j = 0}^{J-1}\left(f_j -s \right)\ket{0}_a\otimes\ket{1}_1\otimes\ket{j}_{n_J}\ .
\end{equation}
This means that we have stored the difference array $f-s$, \emph{i.e.} the array $f$ shifted by the constant $s$, in the second line of the matrix. Note that, at the same time, we have stored the sum array in the first line,
\begin{equation}\label{summ}
    \ket{\phi_3} \supset \ket{\phi_3^{(+)}} \equiv \dfrac{1}{2\|f\|_{\infty}\sqrt{J}}\sum_{j = 0}^{J-1}\left(f_j + s \right)\ket{0}_a\otimes\ket{0}_1\otimes\ket{j}_{n_J}\ .
\end{equation}

\subsection{Approximated linear shift of a given oracle}
\label{lin_shift}

Let us consider a state $\ket{\phi_4}$ obtained by applying a CNOT operator to the state $\ket{\phi_3}$ where the contol qubit is the most significant digit in the register $\ket{j}_{n_J}$ and the target qubit is the row register. This operation flips the row index of the columns corresponding to the second half of the range $\{0,..,J-1\}$, namely those in the range $\{\frac{J}{2},..,J-1\}$. We then have that the $\ket{0}_a\otimes\ket{1}\otimes\ket{j}_{n_J}$ components of $\ket{\phi_4}$ contain the function $f$ shifted by $-s$ in the lower half of points and by $s$ in the higher half, namely
\begin{align}\label{step}
    \ket{\phi_4} \supset \ket{\phi_4^{(-)}} &\equiv \dfrac{1}{2\|f\|_{\infty}\sqrt{J}}\sum_{j = 0}^{\frac{J}{2}-1}\left(f_j -s \right)\ket{0}_a\otimes\ket{1}_1\otimes\ket{j}_{n_J}\\ \nonumber &\qquad + \dfrac{1}{2\|f\|_{\infty}\sqrt{J}}\sum_{j = \frac{J}{2}}^{J-1}\left(f_j +s \right)\ket{0}_a\otimes\ket{1}_1\otimes\ket{j}_{n_J}\ .
\end{align}
This corresponds to a shift by a step function, representing the most rudimentary approximation to a linear function.

So far, we have considered a two row matrix. However, if we want to iterate the step shift just described (for instance to the purpose of getting a finer approximation to a linear shift, we need to consider a larger matrix. Let us start over again with a matrix with $n_I=2$ (that is, $4$ rows) and repeat all the steps that led us to $\ket{\phi_4}$; we add a tilde to indicate that now we are working with a different matrix. We get
\begin{align}\label{step_emb}
    \ket{\tilde\phi_4} \supset \ket{\tilde\phi_4^{(-)}} &\equiv \dfrac{1}{2\|f\|_{\infty}\sqrt{2J}}\sum_{j = 0}^{\frac{J}{2}-1}\left(f_j -s \right)\ket{0}_a\otimes\ket{01}_2\otimes\ket{j}_{n_J}\\ \nonumber &\qquad + \dfrac{1}{2\|f\|_{\infty}\sqrt{2J}}\sum_{j = \frac{J}{2}}^{J-1}\left(f_j +s \right)\ket{0}_a\otimes\ket{01}_2\otimes\ket{j}_{n_J}\ .
\end{align}
Now, let us load a constant array with the value $s'=\frac{2}{3}s$. We can then shift $\ket{\tilde\phi_4}$ by $s'$ and then apply a CNOT operator whose control qubit is the second most significant digit in the column register $\ket{j}_{n_J}$ and whose target qubit is the most significant digit in the row register $\ket{i}_2$. This leads us to eventually choosing the components $\ket{0}_a\otimes\ket{11}_2\otimes\ket{j}_{n_J}$, so we get
\begin{align}
    \ket{\tilde\phi_5} \supset \ket{\tilde\phi_5^{(-)}} &\equiv \dfrac{1}{2\|f\|_{\infty}\sqrt{2J}}\sum_{j = 0}^{\frac{J}{4}-1}\left(f_j - \frac{5}{3} s \right)\ket{0}_a\otimes\ket{11}_2\otimes\ket{j}_{n_J}\\
    \nonumber &\qquad + \dfrac{1}{2\|f\|_{\infty}\sqrt{2J}}\sum_{j = \frac{J}{4}}^{\frac{J}{2}-1}\left(f_j - \frac{1}{3} s \right)\ket{0}_a\otimes\ket{11}_2\otimes\ket{j}_{n_J}\\ \nonumber &\qquad \qquad + \dfrac{1}{2\|f\|_{\infty}\sqrt{2J}}\sum_{j = \frac{J}{2}}^{\frac{3}{4}J-1}\left(f_j + \frac{1}{3} s \right)\ket{0}_a\otimes\ket{11}_2\otimes\ket{j}_{n_J}\\ \nonumber &\qquad \qquad \qquad+ \dfrac{1}{2\|f\|_{\infty}\sqrt{2J}}\sum_{j = \frac{3}{4}J}^{J-1}\left(f_j + \frac{5}{3} s \right)\ket{0}_a\otimes\ket{11}_2\otimes\ket{j}_{n_J}\ .
\end{align}
The process can be iterated, however it becomes inefficient if the step-wise approximation of the linear shift is required to have a precision which, setting the number of discretisation points to $N$, scales as $\frac{1}{N}$.

\section{Discussion and conclusion}
\label{discu}

The main goal of this work is to propose and describe a generic framework for the design of quantum algorithms based on direct embedding. Its modular structure, as depicted in Figure \ref{fig:pipeline}, is appealing and handy in a number of ways. For example, under this framework the main components of a quantum algorithm namely; data loading, arithmetic manipulations, and read-out can be studied and discussed separately.
This holds true also for the considerations related to efficiency, whose current status is reflected into the color coding of Figure \ref{fig:pipeline}; specifically, an end-to-end efficient pipeline would be represented by a left-to-right path within the diagram that encounters only green boxes. Thus, the modular structure of the pipeline for the generic quantum algorithm helps to organise the research effort, compare and interpret different algorithms, and identify possible bottle-necks. Furthermore, it is possible to combine this framework with other existing routines. For instance, it is possible to adopt one's favourite amplitude amplification and estimation technique for the information-extraction part.

On a more technical level, the direct embedding of information into the quantum amplitudes avoids having to deal with square roots and thereby it opens the way to easier arithmetic manipulations of the data stored in the quantum state. In particular, we defined the \emph{quantum matrix}, a two-dimensional array which can be thought of in analogy to a memory register: the basis states correspond to the row and column addresses of the memory locations, while the entries of the matrix are the quantum amplitudes representing the loaded information.
As it has been previously illustrated, this construction allows for neat and flexible manipulation of arrays. We have also covered some basic arithmetic manipulations, for which we provided descriptions and implementation details. All in all, we set up a theoretical proposal for a package of arithmetic operations in a quantum framework. Its full potential and development requires further investigation and work, with particular focus on the loading and read-out modules.

Quantum matrices can be naturally generalized to multi-dimensional arrays. All the arithmetic manipulations proposed, as well as the loading and read-out techniques, can be extended in a  straightforward way to the higher-dimensional and more general tensor setting. However, this comes at the cost of the potential necessity of additional controlled operations needed for ``masking" the array and act only on a desired subset of entries. In other words, the cost of an operation is related to the co-dimension of the subset of entries to which it applies.

Finally, we also provided two specific example applications that are interesting on their own, beyond the discussions of the present work. Namely, the shift of a generic oracle by a constant and the shift by a step-wise approximate linear function. We note that their efficient implementation depends on the efficiency of the oracle to which the shift is applied. A constant shift for an oracle implements a vertical offset and it is useful --for example-- in iterative algorithms where at each iteration an output oracle needs to be centered vertically, \emph{i.e.} along the $y$ axis.

\section{Acknowledgements}
All  authors  acknowledge  the  European  Project  NExt  ApplicationS  of  Quantum  Computing (NEASQC), funded by Horizon 2020 Program inside the call H2020-FETFLAG-2020-01(Grant Agreement 951821).

A. Manzano, Á. Leitao and C. Vázquez wish to acknowledge the support received from the Centro de Investigación de Galicia ``CITIC", funded by Xunta de Galicia and the European Union (European Regional Development Fund- Galicia 2014-2020 Program), by grant ED431G 2019/01.

Part of the computational resources for this project were provided bt the Galician Supercomputing Center (CESGA).

We would like to thank Dario Ferraro, Simon Martiel and Javier Mas for fruitful discussions on some aspects of the present work.

\appendix

\appendix
\section{Details on data loading}
\subsection{Pointwise loading of a matrix}\label{sec:PointwiseLoading}

In this subsection we show how to load a generic matrix (\emph{i.e.} a two dimensional array) into a quantum matrix \eqref{eqn:quantum_matrix_structure} in a pointwise fashion. We will be considering states of the form:
\begin{equation}\label{eqn:quantum_matrix}
    \ket{\hat \psi} = \sum_{i = 0}^{I-1}\sum_{j = 0}^{J-1}c_{ij}\ket{0}_a\otimes\ket{i}_{n_I}\otimes\ket{j}_{n_J}\ ,
\end{equation}
which corresponds to the quantum matrix introduced in \eqref{eqn:quantum_matrix_structure} with the addition of an auxiliary register $\ket{}_a$. In what follows, let us assume for simplicity that the auxiliary register is one-dimensional, \emph{i.e.} it consists of just one qubit. The other registers operate as described earlier when discussing Equation \eqref{eqn:quantum_matrix_structure}.

The pipeline of a quantum algorithm starts by loading an initial state. For example, this can represent a probability distribution and the most simple such case is the uniform distribution. Let us consider it explicitly. To load the uniform distribution, we apply the Walsh-Hadamard gate $\mathbb{1}\otimes H^{\otimes n_I}\otimes H^{\otimes n_J}$ to the base state $\ket{0}_a\otimes\ket{0}_{n_I}\otimes\ket{0}_{n_J}$, thus obtaining:
\begin{equation}\label{eqn:uniform_distribution_loaded}
    \ket{\psi_0} = \dfrac{1}{\sqrt{IJ}}\sum_{i = 0}^{I-1}\sum_{j = 0}^{J-1}\ket{0}_a\otimes\ket{i}_{n_I}\otimes\ket{j}_{n_J}.
\end{equation}
Note that the loading of the distribution has not made use of the auxiliary qubit.

The next step in the pipeline is to load a real matrix $f$ into the quantum matrix. To load a point $f_{ij}$ in the corresponding register we need to act in such a way that we only impact the targeted quantum state. For that purpose we need have to perform three steps:
\\
\begin{wrapfigure}{r}{0.25\textwidth}
    \centering
    \includegraphics[width=0.25\textwidth]{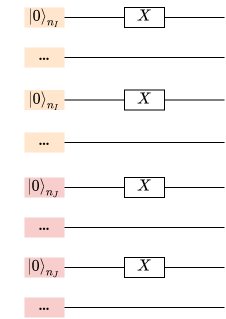}
    \caption{In yellow the row register. In red the column register.}\label{fig:mask}
\end{wrapfigure}
\begin{itemize}
    \item Mask the state. Masking the state consists in converting the state $\ket{0}_a\otimes \ket{i}_{n_I}\otimes \ket{j}_{n_J}$  into the state $\ket{0}_a\otimes\ket{I-1}_{n_I}\otimes \ket{J-1}_{n_J}\equiv \ket{0}_a\otimes\ket{11...11}_{n_I}\otimes \ket{11...11}_{n_J}$. In terms of qubits, this requires to apply a NOT gate to all the qubits that are zero for the original state. The reason for this masking operation can be understood in the next step.
    \item Apply a suitable controlled $y$-rotation on the auxiliary qubit. The controls have to be applied on all the qubits except for the auxiliary one. Here we can see that the complexity of the algorithm depends on the number of qubits that we have to control. 
   The angle for the rotation needs to be $\theta = \arccos\left(\dfrac{f_{ij}}{\|f\|_{\infty}}\right)$, where $\|f\|_{\infty} = \max(|f_{ij}|)$ is the infinity norm of the matrix $f$. The factor $\|f\|_{\infty}$ is needed to keep the amplitudes bounded so that the associated probabilities do not exceed $1$.
    \item Undo step one. This consists in the application of the same mask already used at step one.
\end{itemize}
Following this strategy we can load each of the values $f_{ij}$, thus getting the state:
\begin{equation}
 \begin{split}
    \ket{\psi_1} = &\dfrac{1}{\|f\|_{\infty}\, \sqrt{IJ}}\sum_{j = 0}^{J-1}\sum_{i = 0}^{I-1}\Bigg(f_{ij} \ket{0}_a\otimes\ket{i}_{n_I}\otimes\ket{j}_{n_J}+\sqrt{1-f_{ij}^2} \ket{1}_a\otimes\ket{i}_{n_I}\otimes\ket{j}_{n_J}\Bigg)\ .
 \end{split}
\end{equation}
Usually we focus only on the states marked with $\ket{0}_a$, namely 
\begin{equation}\label{eqn:pointwise_function_loaded}
 \begin{split}
    \ket{\psi_1} \supset &\dfrac{1}{\|f\|_{\infty}\, \sqrt{IJ}}\sum_{j = 0}^{J-1}\sum_{i = 0}^{I-1}f_{ij} \ket{0}_a\otimes\ket{i}_{n_I}\otimes\ket{j}_{n_J}\ .
 \end{split}
\end{equation}

\subsection{Loading a constant array}\label{sec:ConstantArray}

Loading a constant array follows pretty much the same strategy as the pointwise loading. For the purpose of giving an explicit example, we are going to describe the loading of a constant array, taking a real value $c\leq 1$, into a row of the quantum matrix. We start again loading a uniform distribution, thus obtaining \eqref{eqn:uniform_distribution_loaded}. Then, we use a similar structure for loading the array as the one discussed before, namely
\begin{itemize}
    \item Mask the state. In this case we only need to mask the register corresponding to the row (the $\ket{}_{n_I}$ register) and leave the column register untouched. As we mask only one register, we need fewer gates than for the pointwise loading described above.
    \item Apply a suitable controlled $y$-rotation on the auxiliary qubit. The angle for the rotation needs to be $\theta = \arccos\left(c\right)$. The controls have to be made only in the row registers. Here we can see that the number of controls to load a constant array is drastically reduced with respect to the generic function.
    \item Undo step one, by applying the same mask already considered there.
\end{itemize}
If we were to load the constant array in the register $\ket{i}_{n_I}$ we would get:
\begin{equation}\label{eqn:constant_array_loaded}
 \begin{split}
    \ket{\xi_1} = &...+\dfrac{1}{A \sqrt{IJ}}\sum_{j = 0}^{J-1}\Bigg(c \ket{0}_a\otimes\ket{i}_{n_I}\otimes\ket{j}_{n_J}+\sqrt{1-c^2} \ket{1}_a\otimes\ket{i}_{n_I}\otimes\ket{j}_{n_J}\Bigg)+...\ ,
 \end{split}
\end{equation}
where $A$ is a normalization constant. 

As it can be intuitively anticipated, the loading complexity grows together with the lack of symmetry of the loaded state. The constant case, being highly symmetric, is easy. In between the constant and the generic state with no symmetry, one can encounter lower degrees of symmetry, like for example  functions which are piece-wise constant. We remind the reader that we adopted piece-wise constant functions in Subsection \ref{lin_shift} to approximate a linear function and observed how the complexity grew with the approximation accuracy. For further discussions on the relation between the loading complexity and the symmetry of the loaded state we refer to \cite{phdthesis}.

\printbibliography 

\end{document}